\documentclass[aps, twocolumn, pra, showpacs, superscriptaddress]{revtex4}

\usepackage{graphicx}
\usepackage{dcolumn}
\usepackage{bm}
\usepackage{amsmath}

\begin{document}

\title{Fidelity and Quantum phase transition for the Heisenberg chain with the next-nearest-neighbor interaction}

\author{Shu Chen}
%\email{schen@aphy.iphy.ac.cn}
\affiliation{Beijing National
Laboratory for Condensed Matter Physics, Institute of Physics,
Chinese Academy of Sciences, Beijing 100080, China}

\author{Li Wang}
\affiliation{Beijing National Laboratory for Condensed Matter
Physics, Institute of Physics, Chinese Academy of Sciences,
Beijing 100080, China}

\author{Shi-Jian Gu}
\affiliation{Department of Physics and Institute of Theoretical
Physics, The Chinese University of Hong Kong, Hong Kong, China}

\author{Yupeng Wang}
\affiliation{Beijing National Laboratory for Condensed Matter
Physics, Institute of Physics, Chinese Academy of Sciences,
Beijing 100080, China}

\date{\today }

\begin{abstract}
In this paper, we investigate the fidelity for the Heisenberg
chain with the next-nearest-neighbor interaction (or the $J_1-J_2$
model) and analyze its connections with quantum phase transition.
We compute the fidelity between the ground states and find that
the phase transition point of the $J_1-J_2$ model can not be well
characterized by the ground state fidelity for finite-size
systems. Instead, we introduce and calculate the fidelity between
the first excited states.  Our results show that the quantum
transition can be well characterized by the fidelity of the first
excited state even for a small-size system.
\end{abstract}


%05.70.Fh    phase transitions: general studies
%03.67.Mn    entanglement production, characterization, and manipulation
%03.75.Gg    entanglement and decoherence Bose-Einstein condensates
%03.75.Hh    static properties of condensates, thermodynamical statistical
%            and structural properties
%75.10.jm    quantized spin models
%75.30.kz    magnetic phase boundaries (including magnetic transitions,
%            metamagnetism, etc)
%03.67.-a    quantum information
%64.60.-i General studies of phase transitions (see also 63.70.+h
%Statistical mechanics of lattice vibrations and displacive phase transitions;
%for critical phenomena in solid surfaces and interfaces, and in
%magnetism, see 68.35.Rh, and 75.40.-s, respectively)
%75.10.-b General theory and models of magnetic ordering (see also
%05.50.+q Lattice theory and statistics)

%05.70.Fh, 03.67.Mn, 03.75.Gg, 03.75.Hh

% PACS Number

\maketitle
%\section{Introduction}

Quantum phase transitions (QPTs) driven by purely quantum
fluctuations have been extensively studied in the recent years
\cite{Sachdev}. One of the research focuses in the cross field of
quantum many-body theory and quantum-information theory is the
application of quantum entanglement to the analysis of QPTs
\cite{Osterloh,Vidal}. The intriguing issue of the role of quantum
entanglement in characterizing QPTs has been investigated for
different many-body
systems\cite{Osterloh,Vidal,LAmico07,Gu,Gu04,Richter}. More
recently, the ground state fidelity or the overlap between two
ground states corresponding to two slightly different values of
the external parameters is proposed to characterize QPTs
\cite{HTQuan2006,Zanardi06}. Within examples of the Dicke and XY
models, it has been shown that the ground state fidelity shows a
dramatic drop in the vicinity of the QPT point of the system.
Similar to the quantum entanglement, the notation of fidelity is
also borrowed from the field of quantum information science. Being
a pure geometrical quantity, an obvious advantage of the fidelity
is that it can be a promising candidate to characterize the QPT
because no a priori knowledge of the order parameter and the
symmetry of the system is needed \cite{Zanardi06}. By using the
fidelity as a measure, Buonsante {\it et al.} can determine the
quantum phase transition point of the Bose-Hubbard model which is
hard to be characterized by the quantum entanglement
\cite{Buonsante1}.

Despite the success of the ground state fidelity
\cite{HTQuan2006,Zanardi06,Buonsante1,Pzanardi0606130,WLYou07,HQZhou07}
as a measure of QPTs in several concrete examples, it is still not
clear whether the effectiveness of the ground state fidelity in the
study of QPT is general for most of the many-body systems
\cite{WLYou07}. One of the obstacles lies in the difficulty in the
calculation of ground state fidelity because it is generally very
hard to analytically obtain the ground state wavefunction of a
many-body system except a few examples. An even more basic question
is whether the ground state fidelity is a model-independent
indicator for QPTs which exhibits qualitatively different behaviors
at and off the transition point?

In this paper, we will show that the ground state fidelity is not
always a good characterization of the regions of criticality that
define QPTs for a one-dimensional Heisenberg system with
next-nearest-neighbor coupling. Instead, we find that the overlap
of the first excited state or the fidelity of the first excited
state shows a dramatic drop in the vicinity of the QPT point of
the system and can be used to characterize the QPT. We note that
our conclusions are based on the finite size of the chain
considered (up to 24 sites) which may not exclude that the ground
state fidelity for very large systems could in principle be a
characterization of quantum phase transition just like in the case
of XY model \cite{Zanardi06}. However, in the current computation
sources it is not practical to compute a non-integrable spin
system to a very large size as in the case of the exactly solvable
XY model.

The Hamiltonian of a one-dimensional Heisenberg chain with the
next-nearest-neighbor coupling reads
\begin{equation}
 H( \lambda) = \sum_{j=1}^{ L}\left( \hat{s}_j \hat{s}_{j+1} + \lambda \hat{s}_j
\hat{s}_{j+2} \right) , \label{Ham}
\end{equation}
where $\hat{s}_j$ denotes the spin-1/2 operator at the $j\,$th
site, $L$ denotes the total number of sites, and the periodic
boundary conditions $\hat{s}_1= \hat{s}_{L+1}$ are assumed. The
only effective parameter $\lambda$ refers to the ratio between the
next-nearest-neighbor (NNN) coupling and the nearest-neighbor (NN)
coupling. This model is invariant under a global SU(2) rotation,
which implies total spin conservation. For a general $\lambda$,
the model is not analytically solvable. When $\lambda=0$, the
model is exactly solvable by Bethe-ansatz method
\cite{HABethe31,MTakahashib}. When $\lambda=1/2$, the model
reduces to the Majumda-Ghosh model whose ground state is a
uniformly weighted superposition of the two nearest-neighbor
valence bond states \cite{MG}.

The ground-state properties of the model (\ref{Ham}) has been
widely studied by analytical method, such as bosonization and
effective field theory \cite{Haldane,Giamarchi}, and numerical
method, such as exact diagonalization \cite{TTonegawa87,SEgggert}
and density matrix renormalization group
\cite{RJBursill,RChitra,SRWhite96}. The quantum phase transition
driven by the frustration (the competition between the NN and NNN
interaction) is well understood for general $\lambda$. Frustration
due to $\lambda$ is irrelevant when $\lambda<\lambda_c$, and the
system renormalizes to the Heisenberg fixed point, whose ground
state is described as a spin fluid or Luttinger liquid with
massless spinon excitations. As $\lambda>\lambda_c$, the
frustration term is relevant and the ground state flows to the
dimerized phase with a spin gap open. The transition from spin
fluid to dimerized phase is known to be of
Beresinskii-Kosterlitz-Thouless (BKT) type
\cite{Haldane,Giamarchi,Beresinskii,JMKorsterlitz73}. It has been
difficult to determine the BKT point numerically due to the
problem of logarithmic correction \cite{Affleck}. The critical
value of $\lambda_c=0.2411 \pm 0.0001$ has been accurately
determined by numerical methods and conformal field theory method
\cite{Okamoto,Castilla}. The entanglement for the model
(\ref{Ham}) has been studied in Ref. \cite{Gu04} where the ground
concurrences between the nearest-neighbors and the
next-nearest-neighbors are calculated as functions of $\lambda$.
No singularities of the concurrences around $\lambda_c$ are found
for the system with different sizes, which implies that the
concurrences may be not an effective characterization of the QPT.
Very recently, Chhajlany et. al. found that there is a deviation
from the scaling behavior of the entanglement entropy
characterizing the unfrustrated Heisenberg chain when
$\lambda=J_2/J_1 > 0.25$ and thus concluded that this feature can
be used as an indicator of the dimer phase transition
\cite{Richter}.

\begin{figure}[tbp]
\includegraphics[width=9cm]{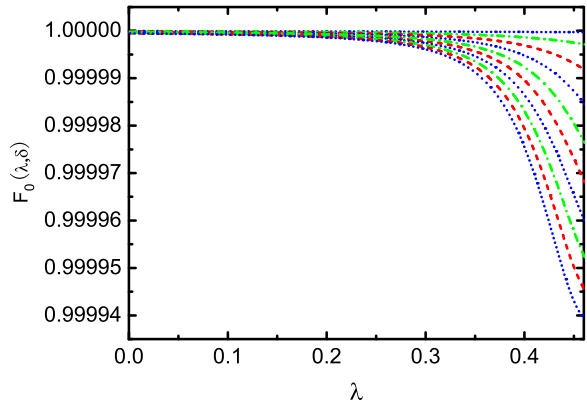}\newline
\caption{(Color online). Ground state fidelity $F_0(\lambda,
\delta)$ as a function of $\lambda$. The lines from top to bottom
correspond to Heisenberg chains with sizes $L= 6, 8, 10, 12, 14, 16,
18, 20, 22, 24$, respectively.} \label{Fig1}
\end{figure}

In the present work, we will study the features of the fidelity for
the model (\ref{Ham}) and focus on the regime of $0<\lambda<0.5$ in
which the BKT-type quantum phase transition happens. Following Ref.
\cite{Zanardi06}, the ground state fidelity is defined as the
overlap between $| \Psi_0(\lambda) \rangle$ and $|
\Psi_0(\lambda+\delta) \rangle$, i.e.
\begin{equation}
F_0(\lambda, \delta) = \left | \langle \Psi_0(\lambda)|
\Psi_0(\lambda+\delta) \rangle \right |,
\end{equation}
where $\Psi_0(\lambda)$ is the ground state wavefunction of
Hamiltonian (\ref{Ham}) corresponding to the parameter $\lambda$ and
$\delta$ is a small quantity. In general, one can numerically solve
the eigenvalue problem of the Hamiltonian and obtain the
eigenfunctions by using the exact diagonalization method for a
finite-size system.

\begin{figure}[tbp]
\includegraphics[width=9cm]{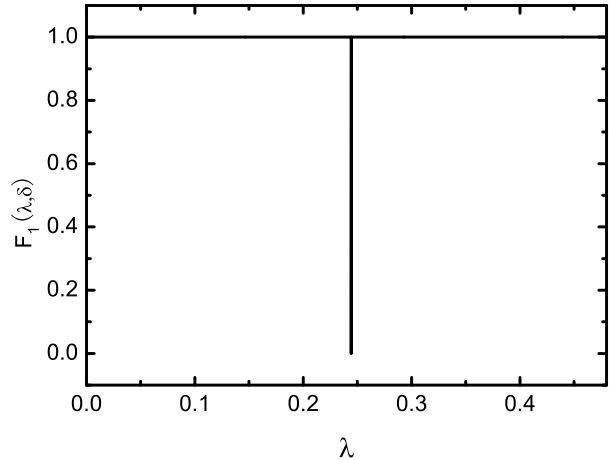}\newline
\caption{The fidelity of the first excited state $F_1(\lambda,
\delta)$ as a function of $\lambda$ for a Heisenberg chain
composed of 10 spin sites.} \label{Fig2}
\end{figure}

We calculate the ground state fidelity of the Heisenberg chain given
by Eq. (\ref{Ham}) for different sizes. In Fig. \ref{Fig1}, we plot
the ground state fidelity as a function of $\lambda$ with
$\delta=1.7\times10^{-3}$ for the frustrated Heisenberg chain with
sizes of $L= 6, 8, 10, 12, 14, 16, 18, 20, 22, 24$. We observe that
the ground state fidelity is almost a constant and equal to unity
for a wide range of the parameter $0<\lambda<0.5$.  According to
\cite{Zanardi06}, one expect a sharp drop of the ground state
fidelity to characterize the critical point of the QPT. However, for
the present model, no a sharp drop in the ground state fidelity is
detected in the regime under investigation for the systems with size
up to 24 sites. Also, we don't find any peaks in the derivatives of
the ground state fidelity, which we do not show here. We note that
no exact analytical results are available for the present $J_1-J_2$
model except the special case of $J_2=0$ and $J_2/J_1=0.5$.
Therefore, we have to calculate the ground state wavefunctions as
well as the ground state fidelity by using the numerical exact
diagonalization method which however limits the size of our
investigated system. Nevertheless, our results imply that critical
points of the quantum phase transitions can not be well
characterized by the ground state fidelity for a finite size system.

%Therefore, we need to find out new tools to reveal the critical
%points.

\begin{figure}[tbp]
\includegraphics[width=9cm]{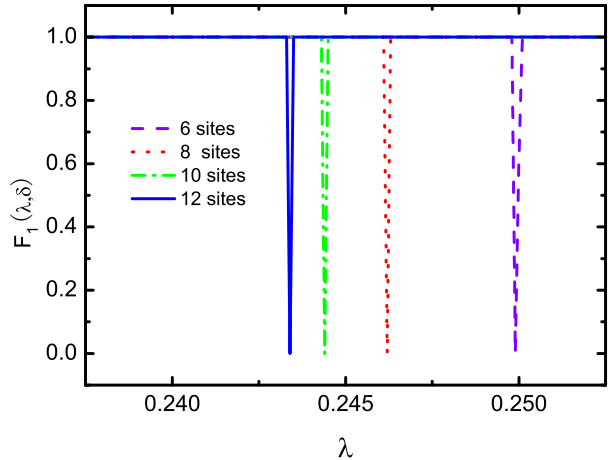}\newline
\caption{(Color online). The fidelity of the first excited state
$F_1(\lambda, \delta)$ as a function of $\lambda$. Different
colors correspond to Heisenberg chains composed of different
numbers of spin sites.} \label{Fig3}
\end{figure}

\begin{figure}[tbp]
\includegraphics[width=9cm]{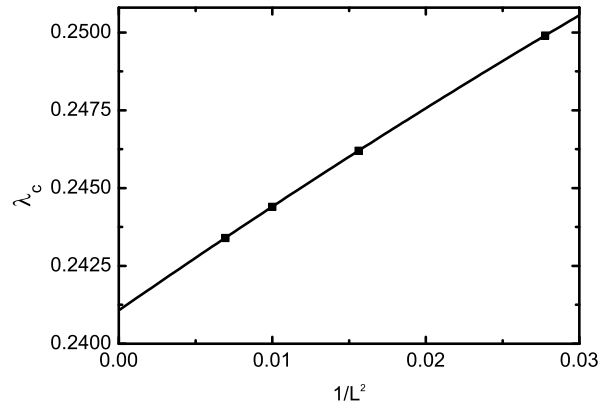}\newline
\caption{Finite size scaling of the extrema of first excited state
fidelity. A polynomial fit is made. According to this fit, when it
comes to the point $L\rightarrow\infty$, $\lambda_c=0.24107$.}
\label{Fig4}
\end{figure}

We recall that, in the scheme of field theory method, the phase
transition point for the model (\ref{Ham}) is determined by the
opening of the elementary excitation gap \cite{Haldane}, which
implies that the excited states play an important role in
determining the phase diagram of the system. Therefore, it is
instructive to investigate the fidelity of the first excited state
of the Heisenberg chain (\ref{Ham}). Similarly, the fidelity of the
first excited state of the system is defined as the overlap of the
first excited states with parameter $\lambda$ and $\lambda+\delta$,
\begin{equation}
F_1(\lambda, \delta) = \left | \langle \Psi_1(\lambda)|
\Psi_1(\lambda+\delta) \rangle \right |.
\end{equation}
where $\Psi_1(\lambda)$ represents the first excited state of the
system. We first calculate the first excited state fidelity of a
Heisenberg chain with $L=10$ as shown in Fig. \ref{Fig2}.
Obviously, there is a sudden drop in the first excited state
fidelity at the point a little smaller than $\lambda=0.25$. This
lights our hope and convinces us that the first excited state
fidelity may be a good candidate to characterize the critical
point between the spin fluid phase and the dimerized phase. From
this point of view, we continue to calculate the first excited
state fidelity of the $J_1-J_2$ model for cases: $L=6, 8, 12$.
Fig. \ref{Fig3} shows the behavior of the first excited state
fidelity $F_1(\lambda, \delta)$  with $\delta=1.7 \times 10^{-4}$
as a function of $\lambda$ for the systems with different sizes.
The extrema of the first excited state fidelity feature a scaling
behavior. The size dependence of the critical point $\lambda_c$
versus $1/L^2$ is shown in Fig. \ref{Fig4}. The four dots
correspond to the four cases: $L=12, 10, 8, 6$. We make a
polynomial fit to the four dots. And we find out that when it
comes to the case: $L\rightarrow\infty$, the critical point is
$\lambda_c=0.24107$. This consists with the value
$\lambda_c=0.2411\pm0.0001$ given by
\cite{RChitra,Okamoto,Castilla} very well. Now we can see that the
ground state fidelity is not always effective for different
models, at least for the $J_1-J_2$ model. For the model considered
in this paper, instead of the ground state fidelity, we need to
rely on the first excited state fidelity to characterize the
critical points of the quantum phase transition.

\begin{figure}[tbp]
\includegraphics[width=9cm]{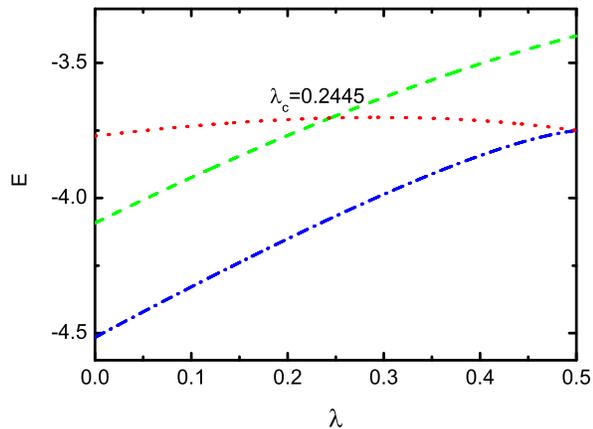}\newline
\caption{ (Color online). The energy spectrum of a Heisenberg
chain with $L=10$. Only the lowest three energy levels are given
above. The first and the second excited energy levels cross each
other at the point $\lambda_c=0.2445$. On the left of $\lambda_c$,
the first excited energy level is triply degenerate and on the
right, it's a singlet. } \label{Fig5}
\end{figure}

To further exemplify why the fidelity of the first excited state
instead of the ground state fidelity is able to characterize the
QPT for the model considered here, let us consider the lowest
energy levels of the model (\ref{Ham}) and analyze its implication
to the fidelity. In Fig. \ref{Fig5}, we plot the energy spectrums
of the Hamiltonian with $L=10$ in the regime of $0<\lambda<0.5$.
The ground state is a singlet with $S_{\rm total}^z =0$ and is
non-degenerate except for the Majumda-Ghosh point with
$\lambda=0.5$. The excited states corresponding to the dashed line
are three-fold degenerate triplet with $S_{\rm total}^z =0,\pm 1$,
whereas the state corresponding to the dotted line is a singlet.
It is clear that no level crossing occurs for the ground state
energy. In general, the first order quantum phase is characterized
by the ground state level crossing which leads to the singularity
of ground state fidelity around to the crossing point. Therefore
the ground state fidelity is a natural choice for characterizing
the first order QPT. When the level crossing of the ground state
is absent, the continuous quantum phase transitions are actually
caused by  a reconstruction (level crossing) of low-excitation
spectrum of the system \cite{Tian}. Therefore for such kind of
system, the fidelity of the first-excited state might be a better
indicator of QPT. The level crossing of the excited state implies
that the corresponding fidelity will suddenly drop to zero in the
crossing point.  This gives a straightforward explanation for why
the fidelity of the excited state is a suitable indicator for the
QPT of the $J_1-J_2$ model.

Though we restrict our attention to the $J_1-J_2$ model, the similar
property can be found in the BKT-like QPTs of other models, such as
the one-dimensional anisotropic Heisenberg model whose Hamiltonian
reads
$$H(\Delta) = \sum_{j=1}^{ L}\left( \hat{s}_j^x \hat{s}^x_{j+1} +
\hat{s}_j^y \hat{s}^y_{j+1} +\Delta \hat{s}_j^z
\hat{s}^z_{j+1}\right).$$ For the anisotropic Heisenberg model, a
BKT-like phase transition happens at the point $\Delta=1$, which
is described by a divergent correlation length but without true
long-range order. However, like the case happened in $J_1-J_2$
model, the fidelity induced by the anisotropic term does not show
the desired singularity at the critical point. This phenomenon is
consistent with the fact that the ground state fidelity
intrinsically depends on the fluctuation of the driving
term\cite{WLYou07}, and such a fluctuation shows no singularity
because of the absence of true long-range order around the
critical point. On the other hand, like $J_1-J_2$ model, the phase
transition in the anisotropic Heisenberg model is also induced by
the first excited state level-crossing \cite{Tian}. This fact
leads to that the first excited state overlap collapses at the
critical point.

In summary, we have calculated the fidelity of the ground state and
the first excited state of the spin chain model with the NNN
interaction. Our results show that, contrary to the first-order QPT
for which the ground state fidelity is a good indicator, the
fidelity of the low-lying excited state is an effective tool to
quantify the quantum phase transition for the system in which the
continuous phase transition is induced by the low-lying excited
states. Though we restrict our calculation on the $J_1-J_2$ model,
our observation is general for a class of BKT-like QPTs, which are
induced by the first excited state level-crossing, in the other
one-dimensional many-body systems, for which the discontinuity of
fidelity of the first excited state is intrinsically related to the
QPTs.

\begin{acknowledgments}
This work is supported by NSF of China under Grant No. 10574150,
programs of Chinese Academy of Sciences, and RGC Grant CUHK 401504.
SJ Gu is grateful for the hospitality of Institute of Physics and
Theoretical Physics at Chinese Academy of Sciences.
\end{acknowledgments}

\end{document}